\documentclass[conference]{IEEEtran}
\IEEEoverridecommandlockouts
\usepackage{amsmath}
\usepackage{amsmath}
\usepackage{amsfonts}
\usepackage{amssymb}
\usepackage{amsthm}
\usepackage{enumerate}
\newcommand{\C}{\mathcal{C}}
\newcommand{\dist}{\mathrm{d}}
\renewcommand{\a}{\alpha}

\newcommand{\G}{\mathcal{G}}
\newcommand{\F}{\mathbb{F}_q}
\newcommand{\Fn}{\mathbb{F}_{q^{n}}}

\newcommand{\n}{\eta}

\newcommand{\rank}{\mathbf{rank}}
\renewcommand{\baselinestretch}{1.05}

\theoremstyle{plain}
\newtheorem{thm}{Theorem}

\newtheorem{pro}{Proposition}

\newtheorem{defn}[pro]{Definition}
\theoremstyle{remark}
\newtheorem{rem}{Remark}

\begin{document}
%
\title{A decoding algorithm for Twisted Gabidulin codes$^{*}$\footnoterule\thanks{$^{*}$This work was supported by SNF grant no. 169510.}}

\author{\IEEEauthorblockN{Tovohery Randrianarisoa}
\IEEEauthorblockA{Institute of Mathematics \\
University of Zurich\\ 
Switzerland\\	
Email: tovohery.randrianarisoa@math.uzh.ch}
\and
\IEEEauthorblockN{Joachim Rosenthal}
\IEEEauthorblockA{Institute of Mathematics \\
University of Zurich\\ 
Switzerland\\	
Email: rosenthal@math.uzh.ch}}


%


\maketitle

\begin{abstract}
In this work, we modify the decoding algorithm for subspace codes by K\"otter and Kschischang to get a decoding algorithm for (generalized) twisted Gabidulin codes. The decoding algorithm we present applies to cases where the code is linear over the base field $\F$ but not linear over $\Fn$.
\end{abstract}


%
\IEEEpeerreviewmaketitle

\section{Introduction}
The first construction of rank metric codes were independently given by \cite{Del78,Gab85,Rot06}. These codes are called Gabidulin codes. For this class of codes, distinct decoding algorithms were provided to correct errors less than half the minimum rank distance. It is also suggested that there are many classes of linear rank metric codes \cite{Ner16} although there are only few known general constructions. In a paper \cite{She15}, Sheekey has provided a new construction of rank metric codes which he called twisted Gabidulin codes.  This class of codes contains the original Gabidulin codes but it also contains codes which are not equivalent to Gabidulin codes. A more general class, called generalized twisted Gabidulin codes were studied in \cite{Lun15}. These rank metric codes have found many applications. For example they can be used in cryptography and recently, they were also used in network coding. But of course, in order to use these codes, decoding algorithms are needed. For Gabidulin codes, as we mentioned, there are already many decoding algorithms in the literature. However, to this time, there is no known algorithm to decode the general class of (generalized) twisted Gabidulin codes. In this work we will provide a decoding algorithm for some special cases of these codes, where they are only linear over $\F$ but not linear over the extension $\Fn$. 

In Section \ref{sec:2}, we will introduce the rank metric codes and the construction of twisted Gabidulin codes. In Section \ref{sec:3}, we will provide the decoding algorithm for the twisted Gabidulin codes. Note that in this work we will consider only codes of full length i.e. with length equal to the degree of the extension $\Fn/\F$ but everything still applies for codes with smaller length.

\section{Rank metric codes}\label{sec:2}
Before we start introducing rank metric codes, we first give the notion of linearized polynomial ring. 
\begin{defn}[Linearized polynomial ring]
Let $\Fn/\F$ be a finite field extension of degree $n$. A linearized polynomial is a polynomial of the form 
\[
f(x)=f_0 x + f_1 x^q + \cdots + f_n x^{q^n}.
\]
If $f_n\neq 0$, then we say that $f(x)$ has $q$-degree $n$. The set of these polynomials, denoted by $L_q[x]$, forms a ring called ``linearized polynomial ring'', where the addition is done componentwise and the multiplication is defined by the non-commutative formula
\[
 a x^{q^i} \circ b x^{q^j}=a b^{q^i}x^{q^{i+j}}, i,j \in \mathbb{N} \text{ and } a,b\in \Fn.
\]
\end{defn}

The above definition of the product is actually the composition of two polynomials.

Since the maps $\Fn\rightarrow\Fn, x\mapsto x^{q^i}$, $i\in \mathbb{N}$ are linear over $\F$, then we see that the linearized polynomials actually define linears maps $\Fn\rightarrow\Fn$ and linearity is over $\F$. That explains the reason why they are used in the context of rank metric codes. The linearity allows us to easily compute the rank norm of a codeword.

By the definition of the product in $L_q[x]$, the ring is non commutative and it is known that a linearized polynomial ring admits a left/right Euclidean division. This division will later be used in the decoding algorithm. For more information on the linearized polynomial ring in the context of coding theory, one can have a look at \cite{Koe08}.

\begin{defn}[Rank metric code]\label{defn:1}
Let $\Fn/\F$ be a field extension. A rank metric $[n,M,d]$-code over $\F$ is a subset of $\Fn^n$ of size $M$ such that the minimum distance between two codewords $x,y$ is $d$. The distance $\dist(x,y)$ is defined by the maximum number of linearly independent elements over $\F$ in the set $\left\lbrace(x_1-y_1),\cdots,(x_n-y_n)\right\rbrace$, where $x=(x_1,\cdots,x_n)$ and $y=(y_1,\cdots,y_n)$.
\end{defn}

The above definition of distance defines a norm which we denote by $\rank(c)$ for a codeword $c$.

\begin{thm}[\cite{Gab85}]
Let $\Fn/\F$ be a field extension and $\C$ be a rank metric $[n,M,d]$-code over $\F$. Then $\log_{q^n}(M)\leq n-d+1$.
\end{thm}

\begin{defn}
Codes for which the inequality in the previous theorem is an equality are called maximal rank distance (MRD) codes.
\end{defn}
Typical examples of rank metric codes were constructed by Gabidulin in \cite{Gab85}. The codes have the parameter $[n,q^{nk},n-k+1]$ and have the property that they are an $\Fn$-subspace of $\Fn^n$. A more general construction was given in \cite{She15}, where it is also possible to get codes which are only $\F$-subspaces. These codes have the property that for given $d,q,n$ and $\Fn/\F$, $M$ is maximal and therefore they are MRD codes. Due to the linearity, the minimum distance of these codes are just the minimum of the $\rank$ norm of the codewords.  We will now give the construction for these codes but before that, we need the following proposition, which is needed for the construction as well as for the decoding algorithm.This is just a reformulation of a theorem in \cite{She15}.

\begin{pro}\label{pro:1}
Suppose that $f$ is a linearized polynomial of $q$-degree $k$ over $\Fn$. If the dimension of the kernel of $f$ in $\Fn$ is equal to $k$, then there is a non-zero $z\in \Fn$ such that $f_0z=(-1)^k f_k^q z^q$.
\end{pro}
\begin{proof}
Suppose that the kernel $V$ of $f$ has dimension $k$. Let $\lbrace x_0,x_1,\cdots,x_{k-1}\rbrace$ be a basis of the kernel of $f$. We define the linearized polynomial $h(x)=\sum_i h_ix^{q^i}$ by
\[
h(x)=\begin{vmatrix}
x & x^q & \cdots & x^{q^k} \\
x_0 & x_0^q & \cdots & x_0^{q^k} \\
\vdots & \vdots & \ddots & \vdots \\
x_{k-1} & x_{k-1}^q & \cdots & x_{k-1}^{q^k} \\
\end{vmatrix}.
\]
It is obvious that $h(x)$ also vanishes on $V$. Therefore, we have that $h_i=f_i z$ for some non-zero $z\in \Fn$. Furthermore, by computing the above determinant, we have $h_0=(-1)^k h_k^q$. The result follows.
\end{proof}

We are now ready to provide the definition of twisted Gabidulin codes which is due to Sheekey \cite{She15}.

\begin{defn}
Let $n,k,r$ be positive integers with $k<n$. Let $\n$ be an element of $\Fn$ such that $N(\n)\neq (-1)^{nk}$. Then the set of linearized polynomials
\[
\G(\n,r)=\lbrace f_0x +f_1x^q+\cdots+f_{k-1}x^{q^{k-1}} +\n f_0^{q^r}x^{q^k} : f_i\in \Fn \rbrace
\]
represents an $\F$-linear MRD code of size $q^{nk}$, which is called a twisted Gabidulin code.
\end{defn}

In the above definition, we have given the rank metric code as a set of linear maps. The rank norm is then the usual definition of the rank of a linear map. This construction was given by Sheekey in his paper \cite{She15}. To see that this construction gives us an MRD codes, we use Proposition \ref{pro:1} to show that the polynomials in $\G(\n,r)$ can have kernel of dimension $k-1$ at most. Namely, since $N(\n)\neq (-1)^{nk}$, we can see that $f_0z\neq (-1)^k \n^q f_0^{q^{r+1}}z^q$ for any $z\in \Fn$.  The rank norm then comes from the rank nullity theorem.

To construct the codewords of the code as we have defined linear rank metric codes in Definition \ref{defn:1}, we actually evaluate these polynomials on a fixed basis $\lbrace \a_1,\cdots, \a_n \rbrace$ of $\Fn$. So our code is in fact
\[
\C=\left\lbrace \left( f(\a_1), f(\a_2), \cdots, f(\a_n)\right), f\in \G(\n,r) \right\rbrace.
\]

\begin{rem}\ 
\begin{enumerate}[(a)]
\item The original Gabidulin codes correspond to the case where $\n=0$.
\item When $r\neq 0$, then we have an MRD code which is linear only over $\F$.
\item This construction can be generalized by replacing the Frobenius map $x^q$ by $x^{q^s}$, where $(n,s)=1$. These are the generalized twisted Gabidulin codes.
\end{enumerate}
\end{rem}

\section{Decoding algorithm}\label{sec:3}
In this section we will provide a decoding algorithm for some twisted Gabidulin codes. The algorithm can be generalized to the case of generalized twisted Gabidulin codes by working on linearized polynomials. The algorithm we present here is a modification of the algorithm by K\"oetter and Kschischang which was used to decode subspace codes \cite{Koe08}.

Let $f\in \G(\n,r)$ be the message polynomial. It is encoded into the codeword 

\[
\left( f(\a_1), f(\a_2), \cdots, f(\a_n) \right).
\]

This codeword is sent and we assume that the word $(r_1,r_2,\cdots,r_n)$ was received. 
We define the error vector
\[
e= \left( f(\a_1)-r_1, f(\a_2)-r_2, \cdots, f(\a_n)-r_n\right)
\]
Under the assumption $t< \frac{n-k+1}{2}$ rank errors happened, we seek the unique codeword $\left( f(\a_1), f(\a_2), \cdots, f(\a_n)\right), f\in \G(\n,r)$ such that $e$ has rank $t$.

$e=(e_1,\cdots,e_n)$ now defines an $\F$-endomorphism 
\begin{align*}
\F^n & \longrightarrow \Fn \\
(b_1,\cdots, b_n) & \longmapsto \sum_i b_i e_i.
\end{align*}
By the definition of $\rank$, we see that the rank of this map is $\rank(e)$. Therefore, by the rank nullity theorem, the kernel $V$ of this map is a subspace of dimension $n-t$.

We have 
\begin{equation}\label{eq:1}
V=\left\lbrace (b_1,\cdots, b_n) : \sum_i b_i f(\a_i) = \sum_i b_i r_i \right\rbrace.
\end{equation}

Assume that we have two linearized polynomials $P_1$ and $P_2$, with degree at most $n-t$ and $n-t-k$ respectively. Assume furthermore that $P_1$ and $P_2$ satisfy
\begin{equation}\label{eq:2}
P_1(\a_i)-P_2(r_i)=0\quad \forall i, 1\leq i\leq n.
\end{equation}
Choose $(b_1,\cdots, b_n)\in V$, we thus have that
\begin{align*}
& P_1(\sum_i b_i \a_i)-P_2(\sum_i b_i r_i) \\=& P_1(\sum_i b_i \a_i)-P_2(\sum_i b_i f(\a_i)) \\
= & P_1(\sum_i b_i \a_i)-P_2( f(\sum_i b_i\a_i)).
\end{align*}
Therefore $P_1-P_2\circ f$ vanishes on a subspace $W\simeq V$ of dimension $n-t$.

First, assume that $P_2$ and $P_1$ are of the form
\begin{align*}
P_1(x)&=a_0x+ a_1x^q+\cdots+a_{n-t-1} x^{q^{n-t-1}} +a_{n-t} x^{q^{n-t}}.
\\
P_2(x)&=b_0x + b_1x^q+\cdots + b_{n-t-k} x^{q^{n-t-k}}.
\end{align*}

and of course 
\[
f(x)=f_0x+ f_1x^q+\cdots+f_{k-1}x^{q^{k-1}}+\n f_0^{q^r} x^{q^{k}}.
\]

With these forms, we see that $P_1-P_2\circ f$ is of $q$-degree $n-t$ at most where the two extreme monomials are
\[
A=(a_0-b_0 f_0) x,
\]
and
\[
B=\left(a_{n-t}-b_{n-t-k}\n^{q^{n-t-k}} f_0^{q^{r+n-t-k}}\right)x^{q^{n-t}}.
\]

%

Now we are ready to explain the decoding algorithm. First, we want to solve the system of linear equations \eqref{eq:2}. This is a system of $n$ independent equations at most. We have $2n-2t-k+2$ unknown. Since we consider that the error has rank, $t<\frac{n-k+1}{2}$, then we see that, the system \eqref{eq:2} is underdetermined. Namely the solution space of this system of equation is of dimension $2$ at least. This can be solved in polynomial time using the Gaussian elimination.

As we have seen, we now have two polynomials $P_1(x)$ and $P_2(x)$ such that $(P_1-P_2\circ f)$ is of $q$-degree $n-t$ but also has $n-t$ zeros. 
If one of $a_0-b_0 f_0$ and $a_{n-t}-b_{n-t-k}\n^{q^{n-t-k}} f_0^{q^{r+n-t-k}}$ is equal to $0$, then we must have $P_1-P_2\circ f=0$ as polynomial. Therefore, we have two possibilities. In the first case we use a division algorithm to recover the polynomial $f$ by computing $P_1/P_2$. Notice that the division is done in the linearized polynomial ring. If that does not work, then the second possibility is that as in the proof of proposition \ref{pro:1}, we must have a polynomial $h(x)=h_0 x + \cdots + h_{n-t}x^{q^{n-t}}$ such that $h$ and $P_1-P_2\circ f$ have the same zeroes and 
\[
h_0=(-1)^{n-t} h_{n-t}^q,
\]
and
\[
h_{n-t}=(-1)^{n-t}\begin{vmatrix}
x_0 & x_0^q & \cdots & x_0^{q^{n-t-1}} \\
x_1 & x_1^q & \cdots & x_1^{q^{n-t-1}} \\
\vdots & \vdots & \ddots & \vdots \\
x_{n-t-1} & x_{n-t-1}^q & \cdots & x_{n-t-1}^{q^{n-t-1}}
\end{vmatrix}
\]

Now the polynomial $P_1-P_2\circ f$ and $h$ differ only by a constant multiple so that, for any solution $\lbrace a_i,b_i\rbrace$ of the system of equations \eqref{eq:2}
\begin{equation}\label{eq3}
\frac{h_0}{h_{n-t}} = \frac{a_0-b_0f_0}{a_{n-t}-b_{n-t-k}\n^{q^{n-t-k}} f_0^{q^{r+n-t-k}}}.
\end{equation}

So we will use the above relation to compute $f_0$ and we will thus get 
\[
\left( g(\a_1^q)-r_1, g(\a_2^q)-r_2, \cdots, g(\a_n^q)-r_n\right).
\]
where 
\[
g(x)=f_1x+\cdots+f_{k-1}x^{q^{k-2}}
\]
Finally, a decoding algorithm of Gabidulin code with allow us to recover $r_i$ and thus we can recover the original message.

So, what remains is how do we find $f_0$ from the relation \ref{eq3}. Suppose we have two linearly independent solutions $\lbrace a_i, b_i\rbrace$ and $\lbrace a'_i, b'_i\rbrace$ of the Equation \eqref{eq:2}. Then they form two polynomials with the same roots as $h(x)$. Thus
\begin{align*}
& \frac{a_0-b_0f_0}{a_{n-t}-b_{n-t-k}\n^{q^{n-t-k}} f_0^{q^{r+n-t-k}}} \\
 = & \frac{a'_0-b'_0f_0}{a'_{n-t}-b'_{n-t-k}\n^{q^{n-t-k}} f_0^{q^{r+n-t-k}}}
\end{align*}
In some case, we can solve this to recover $f_0$. For example, if $r=t+k\mod n$, then this becomes a polynomial equation of degree two. This can be easily solved and we try out the two values of $f_0$ for decoding. This condition is needed here as the above equation cannot be simply solved when the degree is large. Moreover, in that case there may be many possibilities for the solution $f_0$ of the equations and this will render the algorithm impracticable.

We summarize the decoding algorithm in the following. We suppose that $r=t+k\mod n$. And the message polynomial is 
\[
f(x)=f_0x+ f_1x^q+\cdots+f_{k-1}x^{q^{k-1}}+\n f_0^{q^r} x^{q^{k}}.
\]
\subsubsection*{Algorithm}
We are given a received codeword $(r_1,\cdots, r_n)$.

\begin{enumerate}[(I)]
\item\label{step:1} Solve the system of linear equations in the variables $b_j$ and $a_j$,
\[
P_1(\a_i)-P_2(r_i)=0, \quad \forall i, 1\leq i\leq n
\]
where
\[
P_1(x)=a_0 x +\cdots + a_{n-t}x^{q^{n-t}},
\]
and
\[
P_2(x)=b_0x +\cdots + b_{n-t-k}x^{q^{n-t-k}}.
\]
Get two linearly independent solutions $\lbrace a_i,b_i\rbrace , \lbrace a'_i,b'_i\rbrace$.

\item\label{step:2} For the above solutions, compute $P_1/P_2$ as a polynomial division in the linearized polynomial ring. One can easily check if the quotient is the original message by comparing its rank distance to the received codeword.

\item\label{step:3} If the previous step does not give the original message, then solve for $f_0$ in the equation
\[
\frac{a_0-b_0f_0}{a_{n-t}-b_{n-t-k}\n^{q^{n-t-k}} f_0} = 
\frac{a'_0-b'_0f_0}{a'_{n-t}-b'_{n-t-k}\n^{q^{n-t-k}} f_0}
\]
This is equivalent to a polynomial equation of degree $2$.
\item\label{step:4} We now recover $f_0$, and remove the contribution of $f_0x+\n f_0^{q^r} x^{q^{k}}$ from the received word. Any decoding algorithm of Gabidulin codes can now be used to recover 
\[
g(x)=f_1x+\cdots+f_{k-1}x^{q^{k-2}}
\]
\end{enumerate}

For the complexity, we give a brief explanation of why it is polynomial. Step \ref{step:1} is solving a system of linear equations and thus it can be done in polynomial time. In fact finding two linearly independent solutions can be done in $\mathcal{O}(n^3)$. For Step \ref{step:2}, the polynomial division in the linearized polynomial ring as it was described in \cite{Koe08} can also be done in polynomial time. This is $\mathcal{O}(n^2)$ . For Step \ref{step:3}, solving a polynomial system of degree $2$ is easy. Finally, for Step \ref{step:4}, we just need to use any existing decoding algorithm of Gabidulin codes. There are plenty of those and they can be executed in polynomial time \cite{Loi06,Wac13,Rob16}. To conclude, this decoding algorithm is actually dominated by the the first step and thus the overal complexity if $\mathcal{O}(n^3)$. If one uses the the \textsf{Interpolate} procedure in \cite{Koe08}, the overall complexity can even be reduced to $\mathcal{O}(n^2)$. 

\section{Conclusion}
In this paper, we described a decoding algorithm for some particular case of the twisted Gabidulin codes. We have shown that the algorithm can be run in polynomial time. Our algorithm is an adaptation of the decoding algorithm by K\"otter and Kschischang. However, the case only concerns a code which is linear over the field $\F$, where $r=t+k\mod n$. For the practical point of view, we want to decode codes which are linear over the extension $\Fn$, that is for the case where $r=0$. In this regards, more work should be done on this decoding algorithm.The goal of the paper was to introduce an algorithm for decoding twisted Gabidulin codes. This can be tweaked to get faster algorithm but we leave that for a future work.

\bibliographystyle{IEEEtran}
\bibliography{reference}

\end{document}